\documentclass[12pt]{article}\usepackage[hyperfootnotes=false]{hyperref}
\usepackage{epsfig}
\usepackage{float}

\usepackage{caption}

\usepackage{amsmath}
\usepackage{amssymb}
\usepackage{graphicx}
\newlength{\abstractwidth}
\renewcommand{\thefootnote}{\fnsymbol{footnote}}
\renewcommand{\thanks}[1]{\footnote{#1}} 
\newcommand{\starttext}{
\setcounter{footnote}{0}
\renewcommand{\thefootnote}{\arabic{footnote}}}

\newcommand{\be}{\begin{equation}}
\newcommand{\bea}{\begin{eqnarray}}
\newcommand{\eea}{\end{eqnarray}}
\newcommand{\beq}{\begin{equation}}
\newcommand{\ee}{\end{equation}}

\def\eq{&=&}

\def\la{\langle}
\def\ra{\rangle}

\def\simleq{\; \raise0.3ex\hbox{$<$\kern-0.75em
\raise-1.1ex\hbox{$\sim$}}\; }
\def\simgeq{\; \raise0.3ex\hbox{$>$\kern-0.75em
\raise-1.1ex\hbox{$\sim$}}\; }

\def\bi{\begin{itemize}}
\def\ei{\end{itemize}}

\def\CA{{\cal{A}}}

\def\CC{{\cal{C}}}

\def\CA{{\cal{A}}}

\def\Tr{\bf Tr \it}

\def\bn{\bigskip \noindent}

\def\suk{SU(2^K)}
\def\Suk{$SU(2^K)$}

  \def\kl{k-local}
  \def\suk{SU(2^K)}

\makeatletter
\g@addto@macro\normalsize{%
  \setlength\abovedisplayskip{10pt}
  \setlength\belowdisplayskip{20pt}
  \setlength\abovedisplayshortskip{10pt}
  \setlength\belowdisplayshortskip{20pt}
}
\makeatother

\usepackage{color}

\begin{document}
  
\begin{titlepage}

\rightline{}
\bigskip
\bigskip\bigskip\bigskip\bigskip
\bigskip

\centerline{\Large \bf {PiTP Lectures on Complexity and Black Holes }}
\bn

\centerline{\Large \bf {Lecture I }}

\bigskip
\begin{center}
\bf   Leonard Susskind  \rm

\bigskip

 Stanford Institute for Theoretical Physics and Department of Physics, \\
Stanford University,
Stanford, CA 94305-4060, USA \\
\bigskip

\end{center}

\begin{abstract}

This is the first of three PiTP lectures on complexity and its role in black hole physics. 
 
\medskip
\noindent
\end{abstract}

\end{titlepage}

\starttext \baselineskip=17.63pt \setcounter{footnote}{0}
\tableofcontents

\section*{Preface}

\part*{Lecture I: Hilbert Space is Huge}

These lectures are a tale of two metrics on the same space---the space of states of a quantum system of $K$ qubits. One metric is familiar to you and the other probably very unfamiliar. They are extremely different measures of the distance between quantum states and they have different purposes.

The two metrics are totally dissimilar. One, the inner product metric, is small in the sense that no two points are more distant that $\pi/2.$ The other, relative complexity, is huge; almost all points are separated by a distance exponential in $K$. The inner product metric is positively curved like a sphere. The geometry of relative complexity is negatively curved, with the radius of curvature being much smaller than the maximum distance between points (the diameter of the space). Both are compact and homogeneous.

Although they both represent the distance between quantum states they are in a sense incommensurate: The distance between two points can be very small in one metric and exponentially large in the other. They represent very different relations between states.

Relative complexity should be of interest in quantum-computer science but has  not been of much use in the study of ordinary quantum systems, not even black holes if we are interested in the region outside the horizon. It is only when we ask about the interior of a black hole, the region behind the horizon, that relative complexity takes its place as a fundamental quantity.

You might ask why put so much effort into understanding what can never be seen. The answer is two-fold: First Einstein's general theory of relativity predicts and describes the region behind the horizon and if we want to understand quantum gravity we have to take account of this fact. We will not have understood how the full geometry of GR emerges from quantum mechanics without following space into the black hole interior.

The second point is that in a cosmology like ours---de Sitter like---everyone is behind someone else's horizon. We won't understand anything without understanding horizons.

\bn

\section{How Huge?}
Let's consider the space of states of $K$ qubits and make a simple estimate of its size. By size I don't mean the dimensionality of the space defined by the number of mutually orthogonal vectors. I mean something more like the total number of unit vectors. It's of course infinite but we can regulate it.

The dimension of the Hilbert space is $2^K$ and a general vector has the form,

$$|\Psi\ra  = \sum_1^{2^K} \alpha_i |i\ra$$

The $\alpha$ are arbitrary complex numbers. At the moment we won't worry about normalizing $|\Psi\ra$ or dividing out the overall phase. Now let's regulate the infinities by restricting each 
$\alpha_i$ to be one of $m$ values. The total number of states is,

\be
\#states = m^{2^K} = \exp{(2^K \log{m})}
\label{states}
\ee
For $K=4, \ m=4$ the number of states is $4,294,967,296.$

The logarithm of the number of states is more manageable,
\be
\log{(\#states})=2^K \log{m} \approx 22.
\label{logstates}
\ee

There are two interesting things about \ref{logstates}. The first is how strongly it depends on $K,$ namely it grows as $2^K$. The second is how weakly it depends on the regulator parameter $m$. We'll see this trend many times  in what follows.

\section{Volume of CP(N)}

To do any kind of rigorous counting of points in a continuous space  we have to coarse-grain the space. An example is counting states in classical statistical mechanics where we coarse grain, replacing points by little balls of radius $\epsilon.$ This allows us to count states, for example to define entropy,
\be
S=\log{(\#states)}
\label{entropy}
\ee
 at the cost an additive term, $\Delta\log{\epsilon}$, $\Delta$ being the dimension of the phase space. In order to count states in $CP(2^K-1)$ or unitary operators in $\suk$ we have to do something similar.

 The space of normalized states with phase modded out is the projective space $CP{(2^K-1)}$. Let's calculate its volume using the usual Fubini-Study metric. The answer is not hard to guess: it is the volume of the sphere of the same dimension divided by $2\pi.$ The volume of $CP(N)$ is, 
 \be 
V[CP(N)] = \frac{\pi^N}{N!} \approx \left(  \frac{e \pi}{N}   \right)^N
\ee

As I said, in order to count states we need to regulate the geometry. The simplest way to do that is to replace points by small balls of radius $\epsilon.$
The volume of  epsilon-ball of dimension $2N$ is
\be 
V[B(2N)] =\frac{\pi^N}{N!} \epsilon^{2N} \approx  \left(  \frac{e \pi}{N}   \right)^N\epsilon^{2N}.
\ee

The obvious thing to do is to divide the volume of $CP(N) $ by the volume of an epsilon-ball. The result is very simple. The
number of epsilon-balls in $CP(N)$ is  $\epsilon^{2N}.$

Next we replace $N$ by $2^K-1$. We identify the number of states with
 the number of epsilon-balls in $CP(2^K).$ 
\be 
\#states =\left( \frac{1}{\epsilon}\right)^{2(2^K -1)}
\ee
or
\be 
\log{(\#states}) =  2(2^K-1) \log{(1/\epsilon})
\label{balls-in-CP}
\ee

For $K=4$ and $\epsilon = 1/2$ this gives $1,073,741,824$ $\epsilon$-balls.

Equation \ref{balls-in-CP} shows the same pattern as  \ref{logstates}. It grows rapidly with $K$ like $2^K$, and is only logarithmically sensitive to the cutoff parameter.

\section{Relative Complexity}

The enormity of the space of states of quantum systems begs the question of how to measure the distance between states. Are some states close to each other and others enormously far? The usual metric in state space (by which I mean $CP(N)$) is the inner-product metric. It's also called the Fubini-Study (FS) metric. The FS distance between two vectors $|A\ra$ and $|B\ra$ is,
\be 
d_{fs}(A,B) = \arccos{|\la A |B\ra|}.
\ee
It varies from $d_{fs}=0$ when $|A\ra = |B\ra$ to $\pi/2$ when $|A\ra$ and $ |B\ra$ are orthogonal. The space with the FS metric is ``small" in the sense that the furthest distance between points is $\pi/2.$

The FS metric has its purposes but it fails to capture something important. Imagine an intricate piece of machinery with a microscopic switch---a control qubit---that can be flipped on or off. Compare the distance between these two states, 
$$
|\bf{on}\ra     \ \ \   \ \ \   |\bf{off}\ra
$$

 with the distance between two other states: a brand new machine,  and  a pile of rust,
$$
|\bf{new}\ra        \ \ \  \ \ \   |\bf{rust \ pile}\ra
$$

  The FS distance between states $|\bf{new}\ra $ and $|\bf{rust \ pile}\ra$ is exactly the same as the distance between $|\bf{on}\ra $ and $|\bf{off}\ra $, namely $\pi/2$. If the only criterion was the FS distance it would be just as easy to:
%
\begin{enumerate}
\item Make a transition from $|\bf{new}\ra $ to 
$|\bf{rust \ pile}\ra $ as to make a transition from $|\bf{on}\ra $ to $|\bf{off}\ra $.
\item Create a coherent superposition of $|\bf{new}\ra $ and $|\bf{rust \ pile}\ra$, or a coherent superposition of $|\bf{on}\ra $ and $|\bf{off}\ra $.
\item Do a measurement that would be sensitive to the relative phases of such superpositions in the two cases.
\end{enumerate}

But of course this is not true. It is much easier  make a superposition of $|\bf{on}\ra $ and $|\bf{off}\ra $ (it just involves superposing two single-qubit states) than of $|\bf{new}\ra $ and $|\bf{rust \ pile}\ra $. Evidently the FS metric fails to capture the difference between these two cases.

If it were equally possible to physically apply \it any \rm operator to a system, then it would be no harder to superpose the machine and the rust pile, than the on-off states. But in the real world an operator that takes a machine to a rust pile is much more \it complex \rm than an operator that flips a qubit. In practice, and maybe in principle, we carry out very complex operations as sequences of simple operations. Once a set of  ``allowable" simple operations has been identified, a new measure of distance between states becomes possible. It can be defined as:

\bn
\it The minimal number of simple operations needed to go from 
$|A\ra$    to $|B \ra $.\rm  

\bn
In (quantum)information theory the simple operations are called gates and new distance measure is called relative complexity:

\bn \it

The relative complexity of $|A\ra$ and $|B\ra $  is the minimum number of gates required to go from $|A\ra$ to $|B\ra $.
\rm

\bn
One might object that relative complexity is a somewhat subjective idea since it depends on a choice of simple operations. But in a qubit context a simple operation is one that involves a small number of qubits. For example we might allow all one and two-qubit gates  as simple operations. Relative complexity would then be fairly well defined\footnote{There are still details associated with how precisely a product of gates must approximate a target state.} But then a skeptic may say: Yes, but what happens to your definition of complexity if I decide to allow three qubit gates?

I expect (but cannot prove) that the answer is: Not much happens at least for large complexity.  There are two kinds of ambiguities to worry about---additive and multiplicative. The additive ambiguities are associated with precision---how close do we have to get to the target to declare success? If we are required to get within inner-product distance $\epsilon,$ then there is an additive term in the complexity,  logarithmic in $\epsilon.$  It is closely related to the logarithmic ambiguity in  \ref{balls-in-CP} and as we will see in the third lecture, to the $\log{\epsilon}$ that appears in classical entropy.

There is also a multiplicative ambiguity when we go from allowing only one and two qubit gates to allowing three qubit gates. That is because a three qubit gate can be approximated by some number of one and two qubit gates. That ambiguity  can be accounted for as long as the complexity is not too small. 

A basic assumption is that there a set of definitions of quantum complexity that exhibit a universal behavior up to additive and multiplicative ambiguities. 
The important rule is that the allowable gates must be restricted to be \kl \ with $k$ much smaller than the number of qubits comprising the system. All \kl \ means is that the gate involves no more than $k$ qubits.

We'll have more to say about relative complexity but let's first discuss the dual role of unitary operators.

\section{Dual Role of Unitaries}

Complexity can be defined for states\footnote{We continue to work with systems of $K$ qubits. The space of states is $2^K$ dimensional.}, but in many ways it is more natural to start with the complexity of operations that you can do on a system. That means the complexity of unitary operators. Unitary matrices have a dual role in quantum information theory. First of all they are the matrix representation of unitary operators,  
\be 
U = \sum U_{ij}|i\ra \la j|
\ee

Secondly they can be used to represent
maximally entangled states of $2K$ qubits,
\be 
|\Psi\ra = \sum U_{ij}|i\ra |\bar{j}\ra
\ee
In asking about the complexity of unitary operators we are also asking about the complexity of maximally entangled states\footnote{In the case of maximally entangled systems I mean a restricted form of complexity in which gates are only permitted to act on each side of the entangled system separately, thus preserving the maximal entanglement.},

The space of (special ) unitary operators is $\suk.$ It is much bigger than the space of states $CP(2^K-1)$. Let's begin with a crude estimate of its size.
A unitary matrix in $SU(N)$ has $(N^2-1)$ real parameters. If each parameter can take on $m$ values the number of unitary operators is
$$
\#unitaries = m^{(N^2-1)} 
$$
Setting $N= 2^K,$
\bea  
\#unitaries \eq m^{(4^K-1)} \cr \cr
\log{(\#unitaries)} &\approx& 4^K \log{m}
\label{log-Nunit}
\eea
This is the same pattern that we saw for states except that $2^K$ is replaced by $4^K.$

\section{Volume of $SU(2^K)$}

Let's make a more refined calculation of the number of operators in $SU(N)$ by dividing its volume by the volume of an epsilon ball of the same dimensionality (the dimension of $SU(N)$ is $N^2-1.$).
The volume of $SU(N)$ is  (see arXiv:math-ph/0210033) 
\be 
V[SU(N)] = \frac{2\pi^{\frac{(N+2)(N-1)}{2}}}{1!2!3!....(N-1)!}
\ee
The volume of an epsilon-ball of dimension $N^2-1$ is
$$\frac{\pi^\frac{N^2-1}{2}}{  \left( \frac{     N^2-1}{2}  \right)!         }$$

\bn
Using Stirling's formula, and identifying the number of unitary operatos with the number  of epsilon-balls in $SU(N)$ 
\bea   
\#unitaries  &\approx& 
\left( \frac{N}{\epsilon^2} \right)^{\frac{N^2}{2}} \cr \cr
&=&  \left(  \frac{2^K}{\epsilon^2} \right)^{4^K/2}
\label{V-in-e-balls}
\eea
Taking the logarithm,
\be 
\log{(\#unitaries)} \approx  \frac{4^K}{2}K\log{2} + 4^K\log{\frac{1}{\epsilon}}
\label{log-balls}
\ee
which is comparable to \ref{log-Nunit}.
Again, we see the strong exponential dependence on $K$ and the weak logarithmic dependence on $\epsilon.$ The $\log{\frac{1}{\epsilon}}$ term  is multiplied by the dimension of the space.

\section{Exploring $\suk $}

We've seen that the space unitary operators is  gigantic. Now I want to discuss how  to move through it. I've already hinted that we don't make big complexity jumps, but instead move in little steps called gates. A sequence of gates is called a circuit although it has nothing to do with periodicity. It's just a name.

\bn
\bf Definition: \rm

A \kl \ gate is a nontrivial k-qubit unitary operator chosen from some allowed universal gate set. We assume that if $g$ is in allowed set, so is $g^{\dag} $. 

\begin{figure}[H]
\begin{center}
\includegraphics[scale=.3]{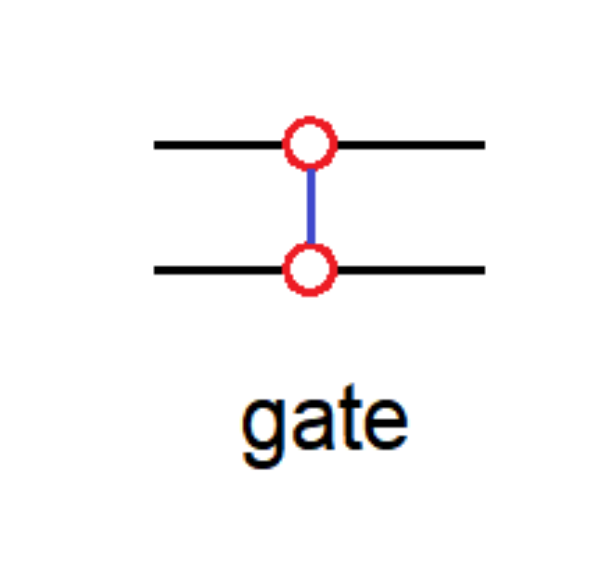}
\caption{Time flows from left to right. A gate acts on an incoming state of two-qubits to give an outgoing state.}
\label{gate}
\end{center}
\end{figure}

Figure \ref{gate} is a schematic representation of a 2-qubit gate.

\begin{figure}[H]
\begin{center}
\includegraphics[scale=.3]{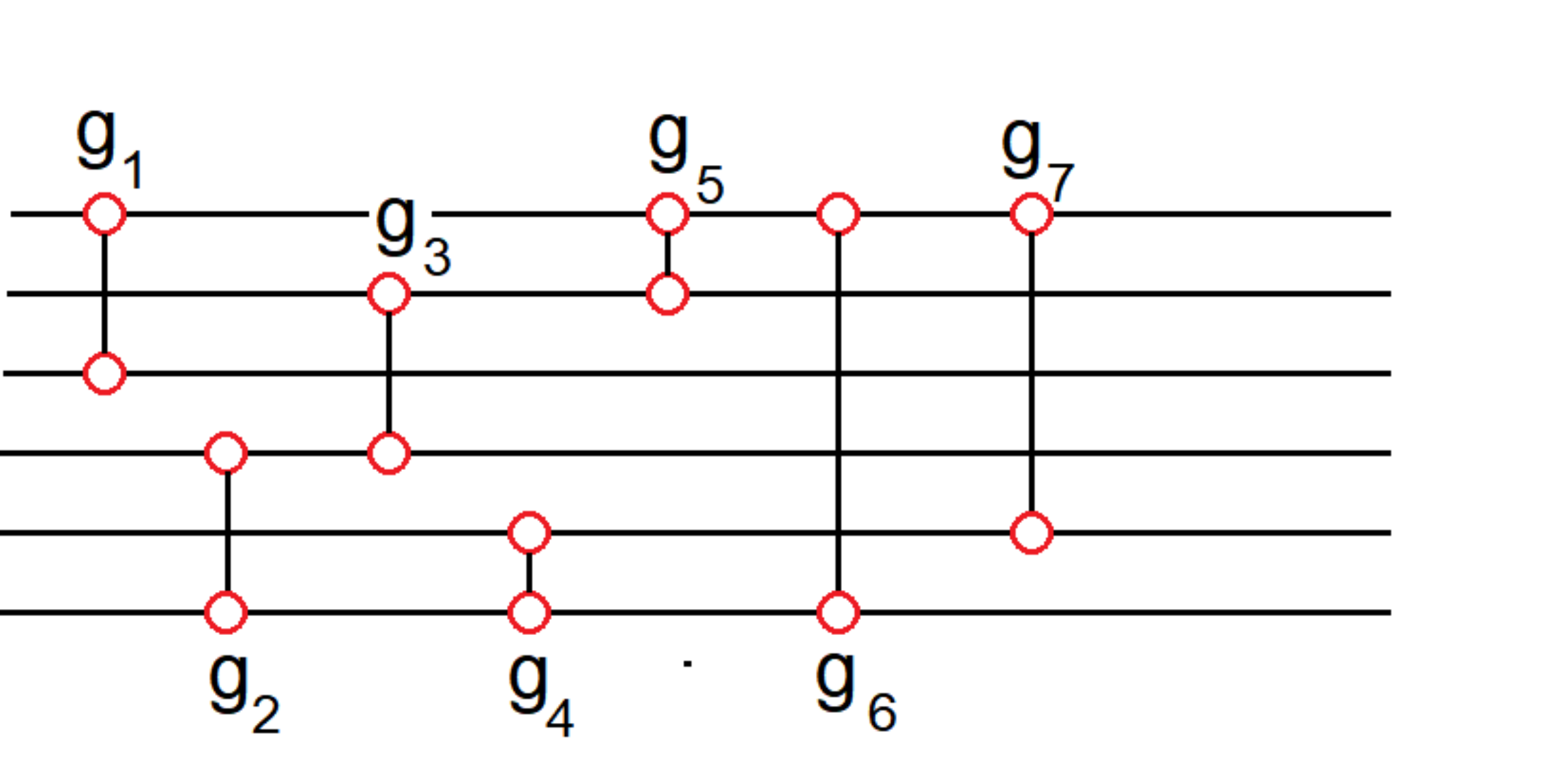}
\caption{k-local all-to-all circuit}
\label{circuit}
\end{center}
\end{figure}

\bn
Gates can be assembled into quantum circuits in the manner shown in figure \ref{circuit}.

\begin{figure}[H]
\begin{center}
\includegraphics[scale=.3]{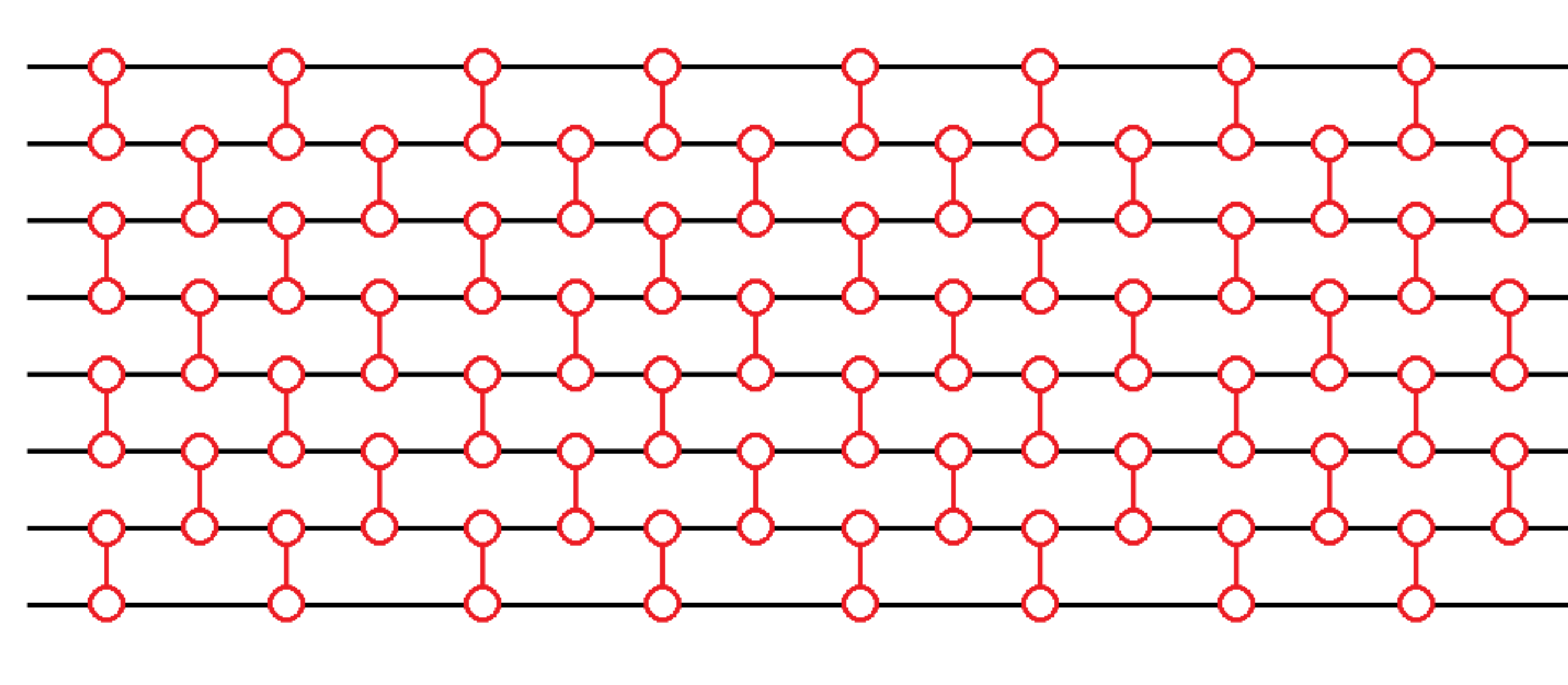}
\caption{Spatially local circuit}
\label{s-local}
\end{center}
\end{figure}

A \kl \ circuit is one made of \kl \ gates. k-locality is not the same as spatial locality although spatially local circuits are special cases of \kl \ circuits. In a spatially local circuit the qubits are arranged on a spatial lattice of some dimensionality. Gates are permitted only between near neigbors on the lattice. An example of a \kl  \ and spatially local circuit is shown in figure \ref{s-local}.
Such circuits would be valuable for simulating condensed matter systems but they are not the kind of circuits we will be interested in.

\bf Definition \rm
A \kl \ \it all-to-all\rm \ circuit is one which is \kl \ but permits any group of $k$ qubits to interact. Figure \ref{circuit} is 2-local \ and all-to-all.

\begin{figure}[H]
\begin{center}
\includegraphics[scale=.3]{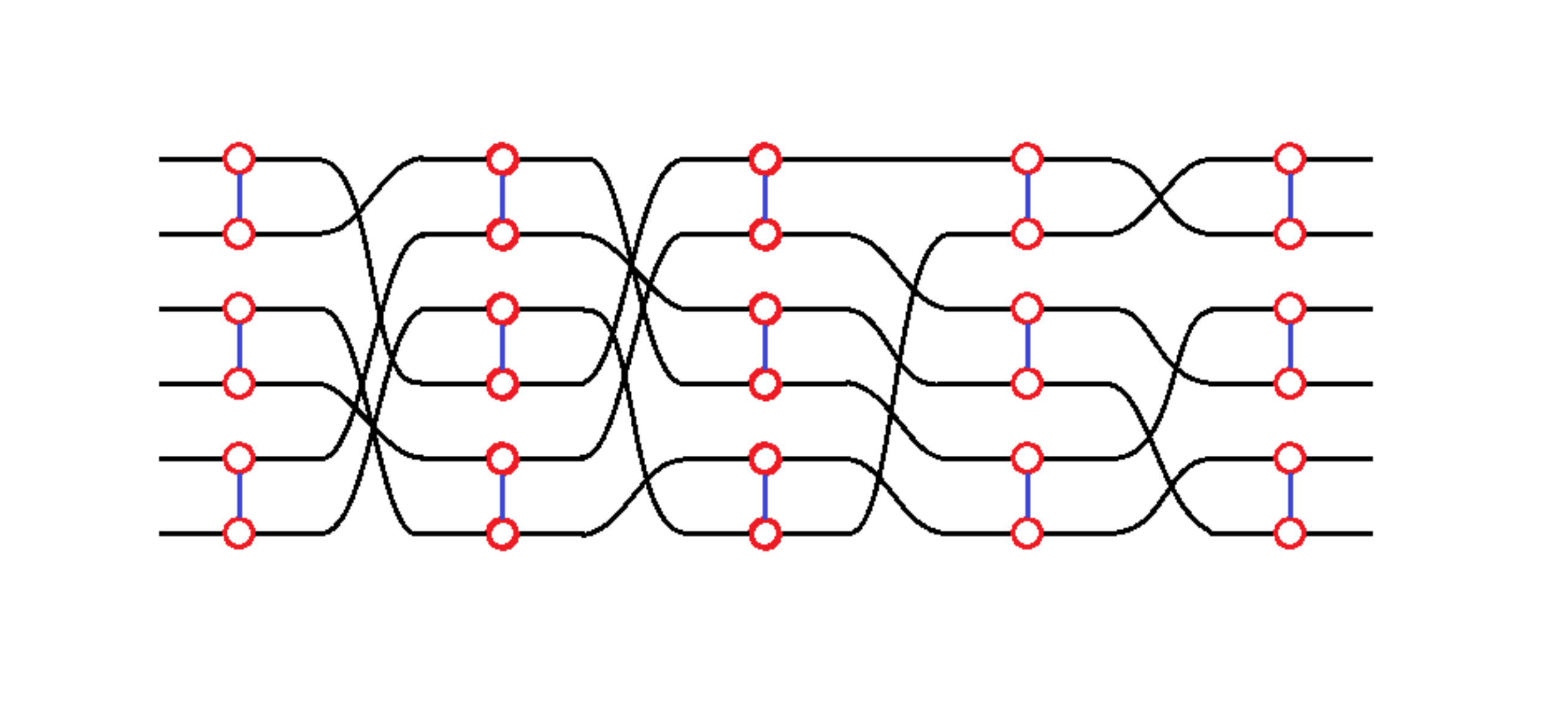}
\caption{Standard circuit architecture. Note that in each step the 
gates all commute because they act on non-overlapping qubits.}
\label{standard-circuit}
\end{center}
\end{figure}

In  figure \ref{circuit} the gates act in series, one gate at a time, but it is more efficient to allow parallel action of gates. The standard \kl \ all-to-all 
circuit architecture is shown in figure \ref{standard-circuit}.
The rule is that at each step the qubits are grouped into pairs and each pair interacts by means of a two-qubit gate.

We imagine that such a circuit is equipped with a clock, and with each tick of the clock $K/2$ gates act. We'll call this a ``step." The number of steps in a circuit is called its depth $D$. The depth of the circuit in figure \ref{standard-circuit} is $D=5.$  The number of gates in the circuit is $\frac{KD}{2}$.

Between steps the qubits may be permuted so that any pair can potentially interact. A circuit of this type can be called \kl \ (2-local in figure \ref{standard-circuit}) and all-to-all. The meaning of all-to-all is that any pair of qubits can potentially interact. Note that in a given time-step the different gates commute because they act on non-overlapping qubit pairs .

If we measure time $\tau$ in units of steps then the number of gates that act per unit time is $K/2.$

\bn

A given circuit prepares a particular unitary operator $U$.
Preparing $U$ by a series of steps can be viewed as a discrete motion of a fictitious classical particle, called the auxiliary system in Brown-Susskind arXiv:1701.01107 [hep-th]. 

The auxiliary system represents the motion of $U(t)$ on $SU(N)$ as time unfolds.  The classical particle  starts at  the identity operator $I$ and ends at $U$.
\begin{figure}[H]
\begin{center}
\includegraphics[scale=.3]{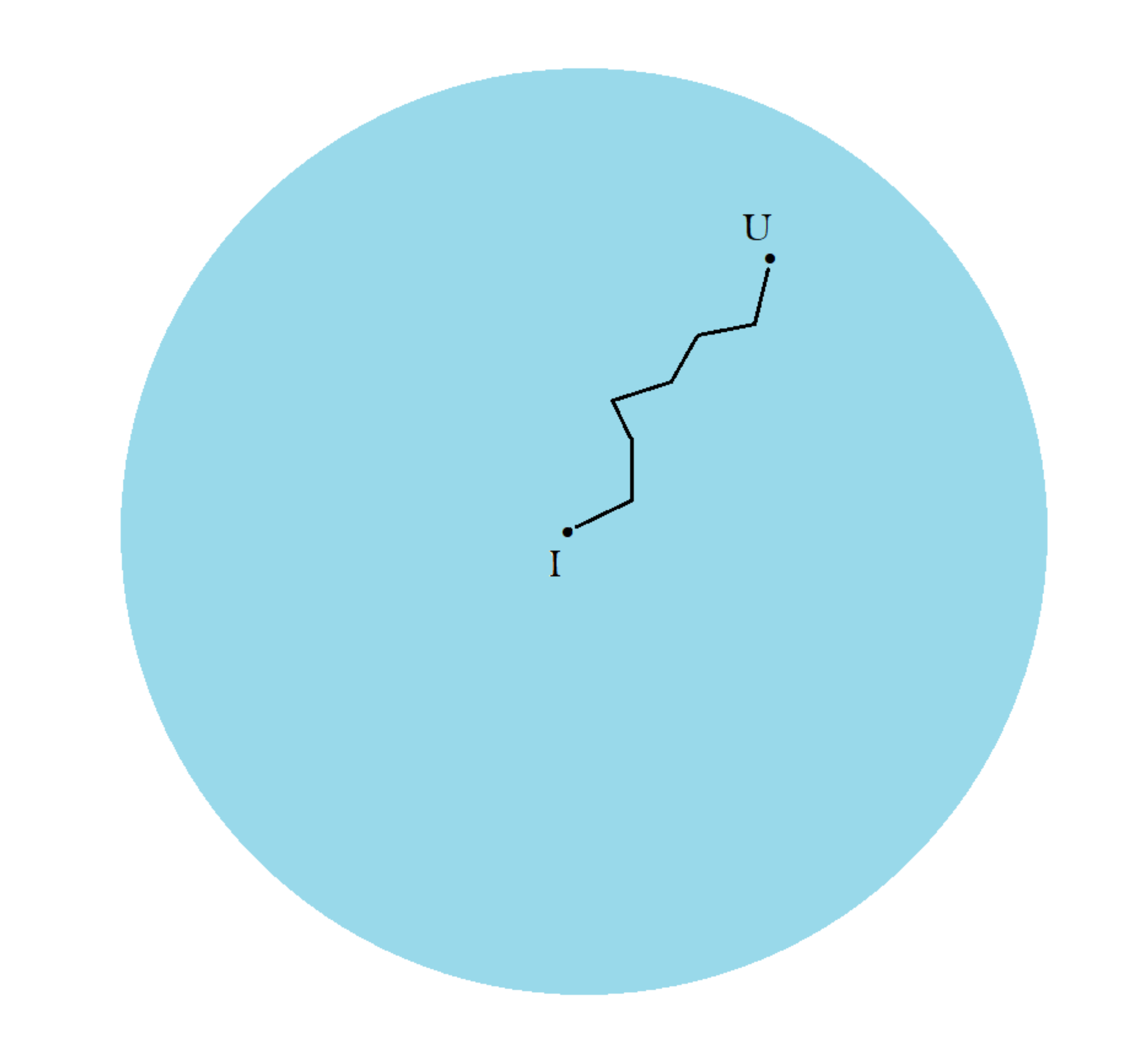}
\caption{The auxiliary system: The evolution of the unitary operator $U(t)$ can be thought of as the motion of a classical particle moving through the group space \Suk.  }
\label{motion}
\end{center}
\end{figure}

\subsection{Relative Complexity of Unitaries}

The standard inner product metric for unitaries is similar to the inner product metric for states. In fact if we think of unitaries as the wave functions of maximally entangled systems then it is the inner product for such states. The inner product distance between $U$ and $V$ is,
\be 
d(UV) = \arccos \   |\Tr U^{\dag}V|
\ee
where $\Tr$ means the normalized trace defined so that $\Tr I =1.$
For the same reasons that I discussed earlier, the inner product distance is not a useful measure of how difficult it is to go from $U$ to $V$ by small steps. A much better measure is relative complexity.\\

\bn
Given two  unitaries the relative complexity $\CC(U,V)$ is defined  as the minimum number of gates (in the allowed gate set)  satisfying,
\be 
U = g_n g_{n-1}....g_1 V
\label{U=ggggV}
\ee
to within tolerance $\epsilon.$ The relative complexity of $U$ and the identity may be defined to be the complexity of $U$.
\be 
\CC(U) \equiv \CC(U,I).
\ee

\begin{figure}[H]
\begin{center}
\includegraphics[scale=.3]{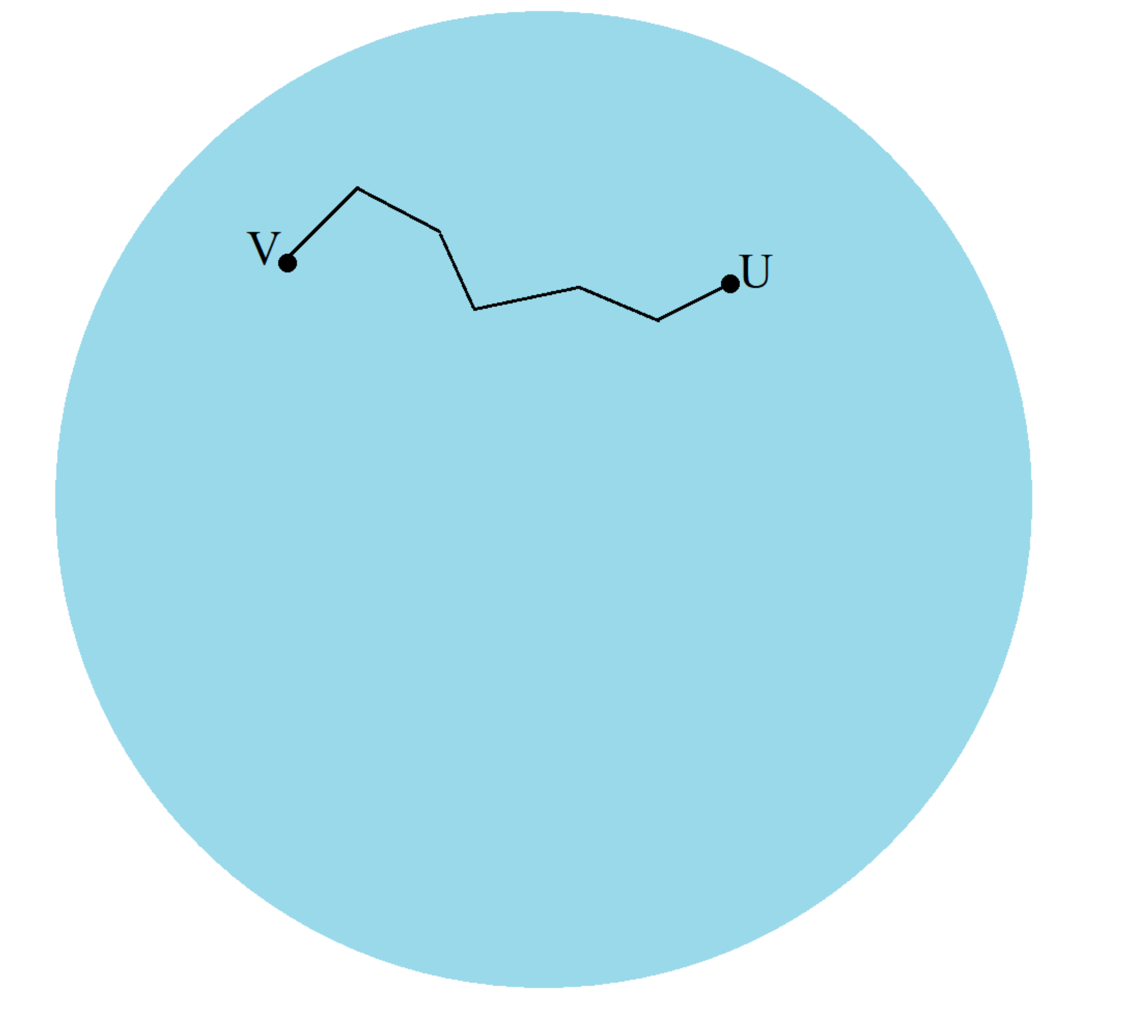}
\caption{Relative-C}
\label{relative-C}
\end{center}
\end{figure}

Figure \ref{relative-C} shows how the relative complexity can be thought of in terms of a discrete curve from $V$ to $U$.

\bn
Because the curve defining $\CC(U,V)$ is the shortest such path, we can think of it as a geodesic, but NOT a geodesic with respect to the inner product metric. It is a geodesic with respect to  relative complexity. The geometry defined by relative complexity is very different from the geometry defined by the inner product distance.

\bn
\bf Relative complexity is a metric \rm

\bn

Relative complexity satisfies the four defining conditions to be a metric.

\begin{enumerate}
\item $\CC \geq 0.$
\item $\CC(U,V)= 0 \ \rm iff \it \ U=V $
\item  $\CC(U,V) = \CC(V,U)$
\item  $\CC(U,V) \leq \CC(V,W) + \CC(W,V)$ \ \ \ \rm (Triangle \ Inequality)
\end{enumerate}

In fact it is a particular kind of metric called  \it right-invariant.\rm  

\bn
Suppose that 
\be 
U=(g_n g_{n-1}.......g_1) \  V
\label{U=ggggV2}
\ee
Then for any $W$ it follows that,
\be 
UW=( g_n g_{n-1}.......g_1) \  VW.
\label{U=ggggV3}
\ee
In other words the relative complexity of $U$ and $V$ is the same as that of $UW$ and $VW.$ 
\be 
\CC(U,V) = \CC(UW,VW) \ \ \ \ \ \ \rm (all  \it W)
\ee
This is what it means for $\CC$ to be  right-invariant.

\bn
On the other hand if we multiply from the left,
\be
WU=(W g_n g_{n-1}.......g_1W^{\dag}) \  WV.
\label{U=ggggV4}
\ee
But $(W g_n g_{n-1}.......g_1W^{\dag})$ 
generally is not a product of $n$ allowed gates. Therefore $\CC$ is not left-invariant\footnote{The usual inner-product metric is both left and right invariant. It is called bi-invariant.}.


\bn
It should be obvious from these remarks that quantum complexity is really a branch of geometry---right-invariant geometry\footnote{See Dowling, Nielsen.}

Most of the mathematical literature on the subject is about left-invariant geometry but of course this is just a matter of convention.

\subsection{Complexity is Discontinuous}
\bn
In describing relative complexity  I've  suppressed issues having to do with coarse-graining by epsilon balls. For example in \ref{U=ggggV}, \ref{U=ggggV2},  \ref{U=ggggV3}, and \ref{U=ggggV4}
I  ignored the phrase, \it to within a tolerance epsilon.\rm \ In fact as epsilon becomes small it takes an ever increasing number of gates to achieve that tolerance. The increase is only logarithmic in $\frac{1}{\epsilon}$ but nevertheless, with the present definition there is not a formal limit of relative complexity, or of the geometry of relative complexity.  One can only define a sequence of geometries as $\epsilon $ decreases.

The relation between the familiar inner-product metric and the relative complexity metric becomes wildly discontinuous as $\epsilon\to 0.$ For example two points can be close in complexity space, e.g., $|\bf {on}\ra$ and $|\bf {off}\ra$, and be maximally distant in the inner product metric. Similarly two states can be close in inner product and far in complexity. When this happens the two states have all expectation values close to one another, which means they yeild almost identical results for any experiment. Nevertheless 
 making a transition between them requires many gates.

It is an interesting question why a physicist would ever be interested in  distinctions which have essentially no effect on expectation values. The answer is that if it were not for black holes, most likely no physicist would be.

\section{Graph Theory Perspective}

Circuits can be usefully described using graph theory\footnote{See also Henry Lin, ``Caley Graphs and Complexity Geometry",  arXiv:1808.06620 [hep-th].}. I'll call a graph  describing a circuit a \it circuit graph.\rm

Let's begin at the identity and act with a circuit of depth one, in other words a circuit of a single step with $K/2$ gates. For simplicity let's assume the allowed gate set is a single non-symmetric two-qubit gate. A choice must be made of how the $K$ qubits are paired. Each pairing will lead to a different unitary. I'll call the number of possible choices $d$ (an odd choice of notation but it corresponds with standard graph-theory notation). It's an easy combinatoric exercise to compute $d$,
\be 
d \sim \frac{K!}{\left(  K/2 \right)!}  \sim \left(  \frac{2K}{e}   \right)^{\frac{K}{2}}
\ee

Let's visualize this by a decision tree embedded in $\suk$. At the central vertex is the identity operator. The branches (or edges) correspond to the $d$ choices. 

\begin{figure}[H]
\begin{center}
\includegraphics[scale=.3]{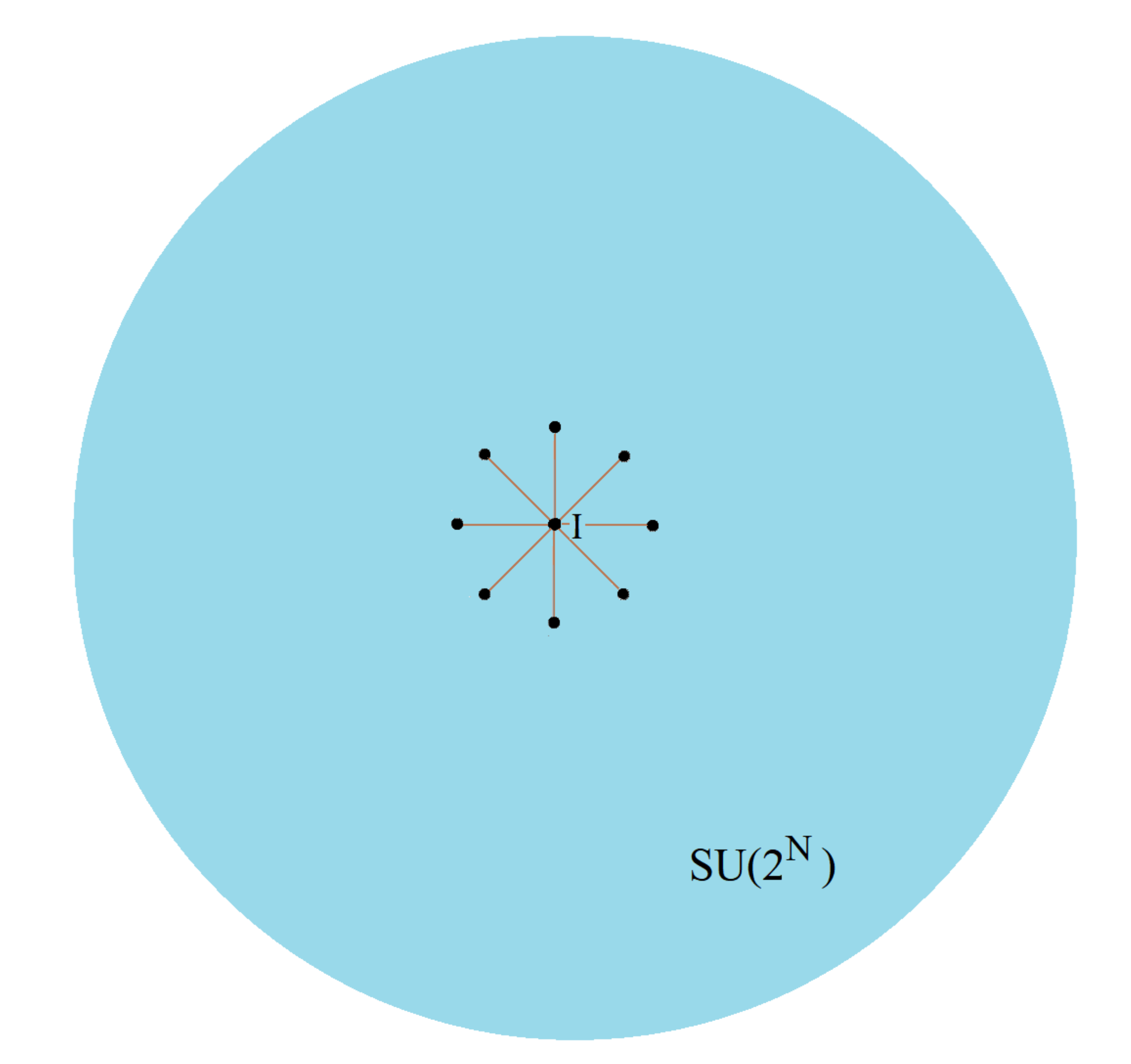}
\caption{Vertex with $d=8$.}
\label{spider}
\end{center}
\end{figure}
The number of branches or edges $d$ at a vertex is called the degree of that vertex. In our construction every internal vertex has degree $\frac{K!}{\left(  K/2 \right)!}$. A graph in which every vertex has degree $d$ is called d-regular. But so far the circuit graph is not d-regular because the outer boundary vertices have a single edge.
The leaves of the graph are unitary operators in $\suk.$

\bn

\begin{figure}[H]
\begin{center}
\includegraphics[scale=.3]{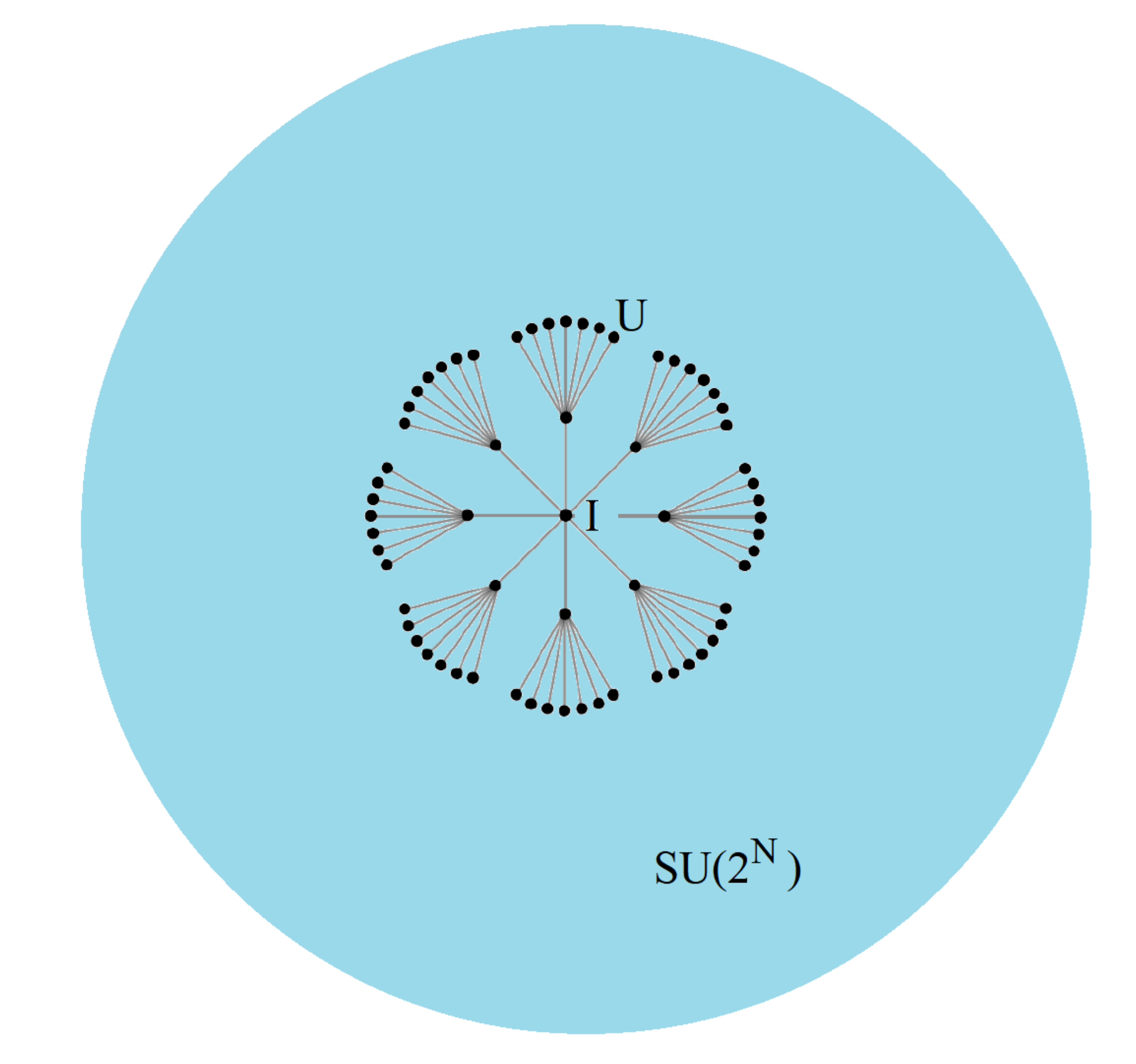}
\caption{Regular tree with degree 8, depth 2.}
\label{ssspider}
\end{center}
\end{figure}

Now let's grow the circuit depth by adding another step. I'm going to make one restriction of a technical nature. For large $K$ it makes no difference but it simplifies things. I will assume that when I add a step the choice of pairings is not the same as in the previous step. This  implies that the next layer of the tree has 
$$\frac{K!}{\left(  K/2 \right)!}-1$$
new branches. This is shown in figure \ref{ssspider}
Now comes an important assumption which underlies everything I will say. Each endpoint (leaf) of the tree represents a unitary operator. I will assume that the probability of two leaves being the same unitary is vanishingly small. Basically the reason is the extremely high dimension of the space of unitaries. If two leaves represent the same unitary to within $\epsilon$ I will say that a collision occurred. My provisional assumption is that collisions are so rare that they can be ignored.

This no-collision assumption  must eventually break down as the tree grows outward and we will calculate when that happens. But for now let's ignore collisions.

Assuming no-collisions, the number of unitaries that have been reached at depth $D$  is,
\be 
\#unitaries=d^D \approx \left(  \frac{2K}{e}   \right)^{\frac{DK}{2}}
\ee

The number of gates in a circuit of depth $D$ is $DK/2$. Assuming 
no-collisions, the path to each leaf is the minimal path, implying that the number of gates is the complexity. Thus we can write,

\bea 
\#unitaries \it \eq \left(  \frac{2K}{e}   \right)^{\CC} \cr \cr
 &\sim & e^{\CC \log{K}}
 \label{N=expC}
 \eea
Let me rephrase this formula in the following way:

\bn
\it The sub-volume of $\suk$ that corresponds to unitaries of complexity $\CC$ grows exponentially with $\CC.$ \rm

\bn
It says a number of things. First of all if we think of complexity as the distance from the origin (the unit operator ) then it says that the volume grows exponentially with radius---a characteristic of hyperbolic spaces of negative curvature. This may seem surprising: with the usual bi-invariant metric $\suk$ is positively curved. But we are talking about a different metric, relative complexity. This negative curvature is a symptom of chaos and in my opinion it, not fast scrambling, is the general signature of quantum chaos.

Equation \ref{N=expC} suggests something else. In classical statistical mechanics the volume of phase space associated with states of a given entropy is exponential in the entropy. The exponential growth of volume in $\suk$ associated with a given complexity is the basis for a deep analogy between complexity and entropy including a \it Second Law of Complexity. \rm It can be summarized by the slogan:  Complexity is the entropy of the auxiliary system $\CA.$

\bn

\begin{figure}[H]
\begin{center}
\includegraphics[scale=.2]{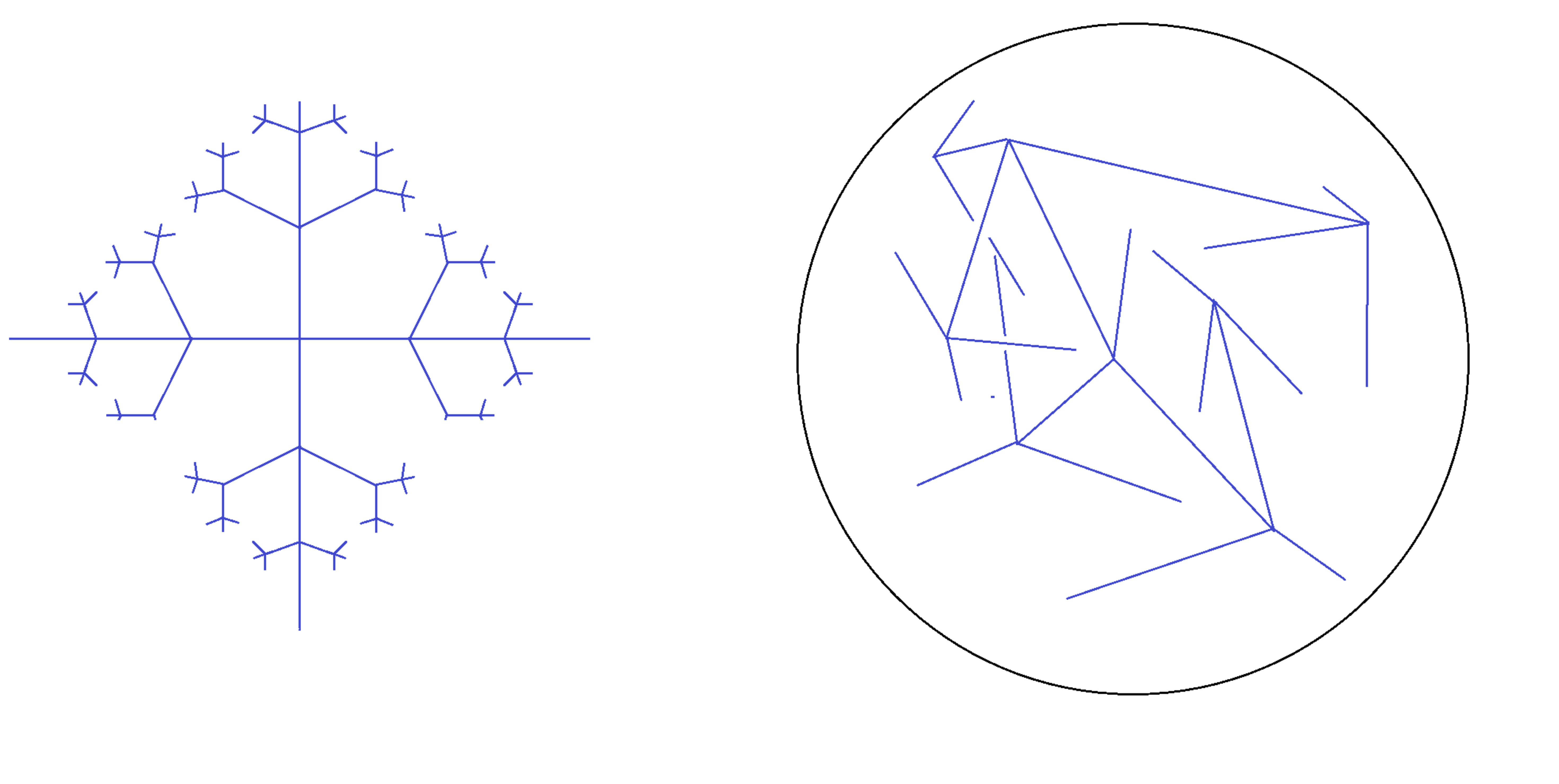}
\caption{Two views of the circuit tree. On the left is an abstract decision tree. The complexity of the unitary leaf grows linearly  with the number of steps from the identity at the center. Relative complexity is proportional to the number of steps between leaves. On the right the tree is embedded in \Suk \ with the usual inner product metric.}
\label{two-views}
\end{center}
\end{figure}

One more point about the circuit graph and its embedding in $SU(2^K)$. Figures \ref{spider} and \ref{ssspider} are very schematic. It is difficult to accurately convey the actual properties of the embedding with pictures given the fact that the space is extremely high dimensional. One thing to keep in mind is that with the usual bi-invariant metric, \Suk \ is small in the sense that the largest distance is $\pi/2.$ A single one-step circuit will move $U$ about that distance. So although the tree abstractly grows outward from the center as in the left panel of figure \ref{two-views}, the embedding in \Suk \ looks more like the right panel. 
The epsilon balls get filled in by the growing circuit in a very fractal manner which is related to the fact mentioned earlier that complexity is discontinuous as $\epsilon\to 0.$

\bn

\subsection{Collisions and Loops}

\begin{figure}[H]
\begin{center}
\includegraphics[scale=.4]{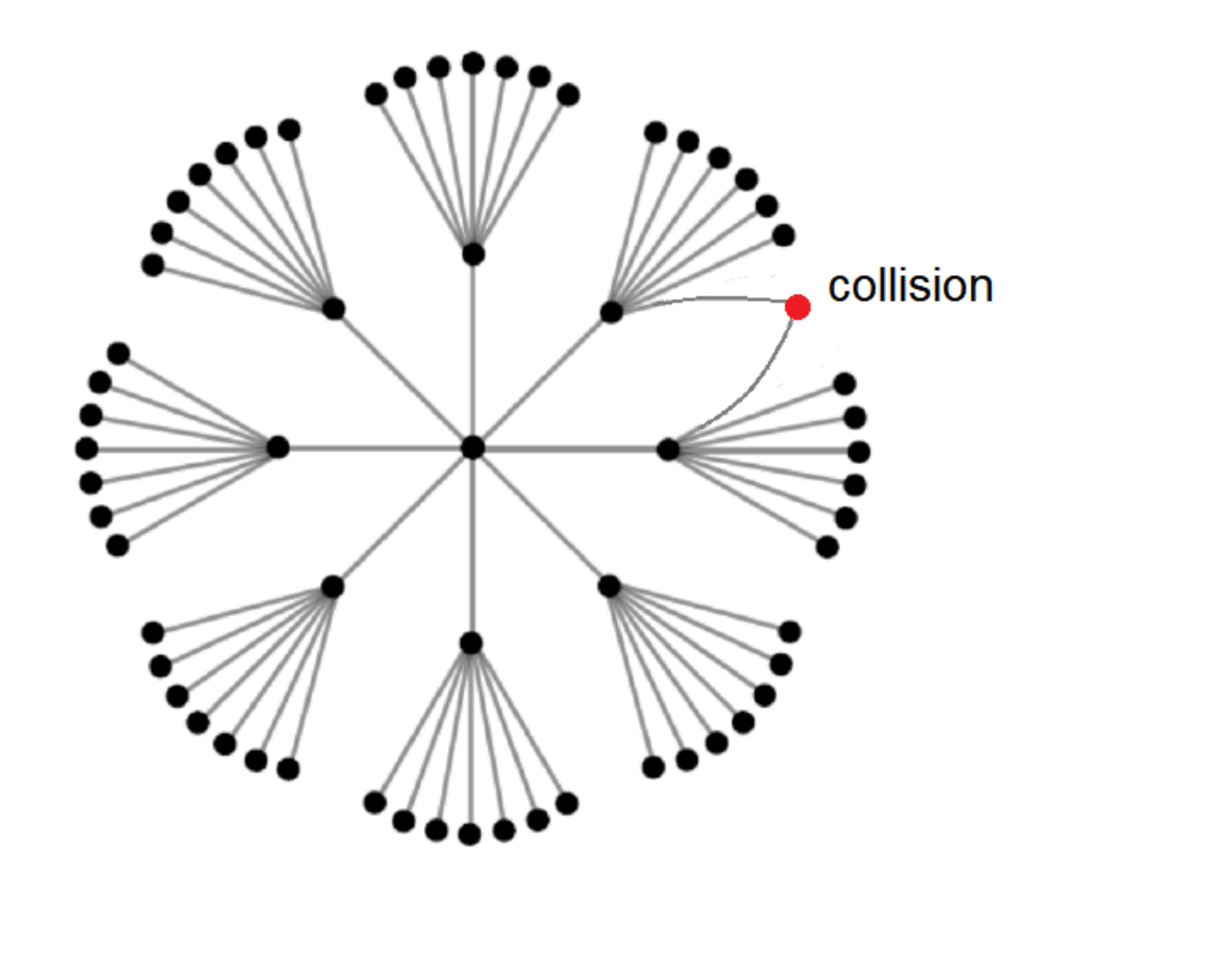}
\caption{Two paths collide on the same epsilon-ball.}
\label{collision}
\end{center}
\end{figure}

Each vertex of the tree represents a unitary operator and for that reason we can think of the tree as being embedded in $SU(2^K).$ 
Suppose a collision does occur. This means that two leaves of the tree are located at the same point is $\suk.$ We can represent this by drawing the two leaves as a single leaf as in figure \ref{collision}. 
The figure illustrates the fact that collisions induce loops in the graph. This allows us to translate the rarity of collisions into graph-theoretic terms. No collisions at all would mean no loops, i.e., exact tree-ness. But as we will see, eventually for very large depth, collisions must happen. This means that very large loops must occur. The correct formulation, which we will come back to, is that small loops must be absent or very rare. 

\bn

 Since the number of epsilon-regulated unitaries is finite we will eventually
 run out of room on $\suk$.  That happens when the number of leaves (as given by \ref{N=expC}) is equal to the total number of epsilon-balls as given in
  \ref{V-in-e-balls}. This determines the maximum possible complexity.

\be 
\left(  \frac{2K}{e}   \right)^{\CC_{max}} = \left(  \frac{2^K}{\epsilon^2} \right)^{4^K/2}
\ee
or

\be 
\CC_{max} = 4^K \left[\frac{1}{2} +\frac{|\log{\epsilon}|}{\log{K}}
\right]
\ee
Again, strong dependence on $K$, weak dependence on $\epsilon.$
\bn
Roughly  $$\CC_{max} \sim 4^K.$$

Apart from a factor of $K$ which is swamped by the exponential, this is also the depth at which collisions must occur. In other words it is the maximum radius at which the tree stops being tree-like. Finally it is also the largest distance  between nodes of the tree---the diameter of the graph.

We can now state the no-collision assumption more precisely. 

\bn

\it 

Loops smaller than $4^K$ are absent or very rare.\rm  

\bn
In graph-theoretic terms, the girth of the graph is $\sim 4^K.$

\bn
Another point follows from the fact that the total number of unitaries is $\sim   e^{4^K}$. We may identify this with the number of vertices in the graph. This implies that the diameter of the graph is logarithmic in the number of vertices.

\bn

\subsubsection*{The breakdown of no-collisions}
\begin{figure}[H]
\begin{center}
\includegraphics[scale=.3]{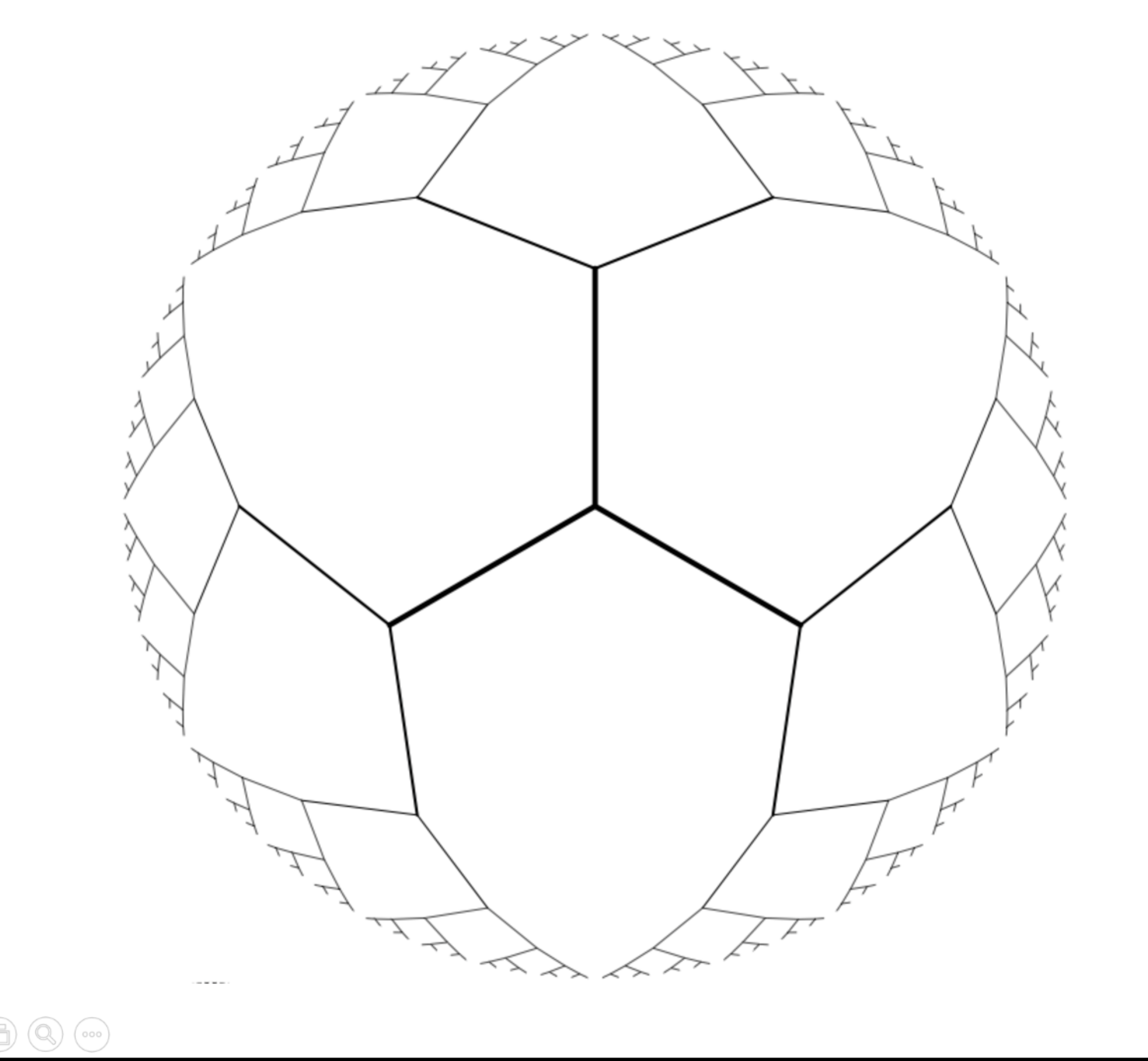}
\caption{Regular tree with degree 3}
\label{bethe}
\end{center}
\end{figure}

Let's consider what happens when the tree-ness breaks down.
Up to that point the graph is a d-regular tree similar to figure \ref{bethe} except with much higher degree.

\begin{figure}[H]
\begin{center}
\includegraphics[scale=.3]{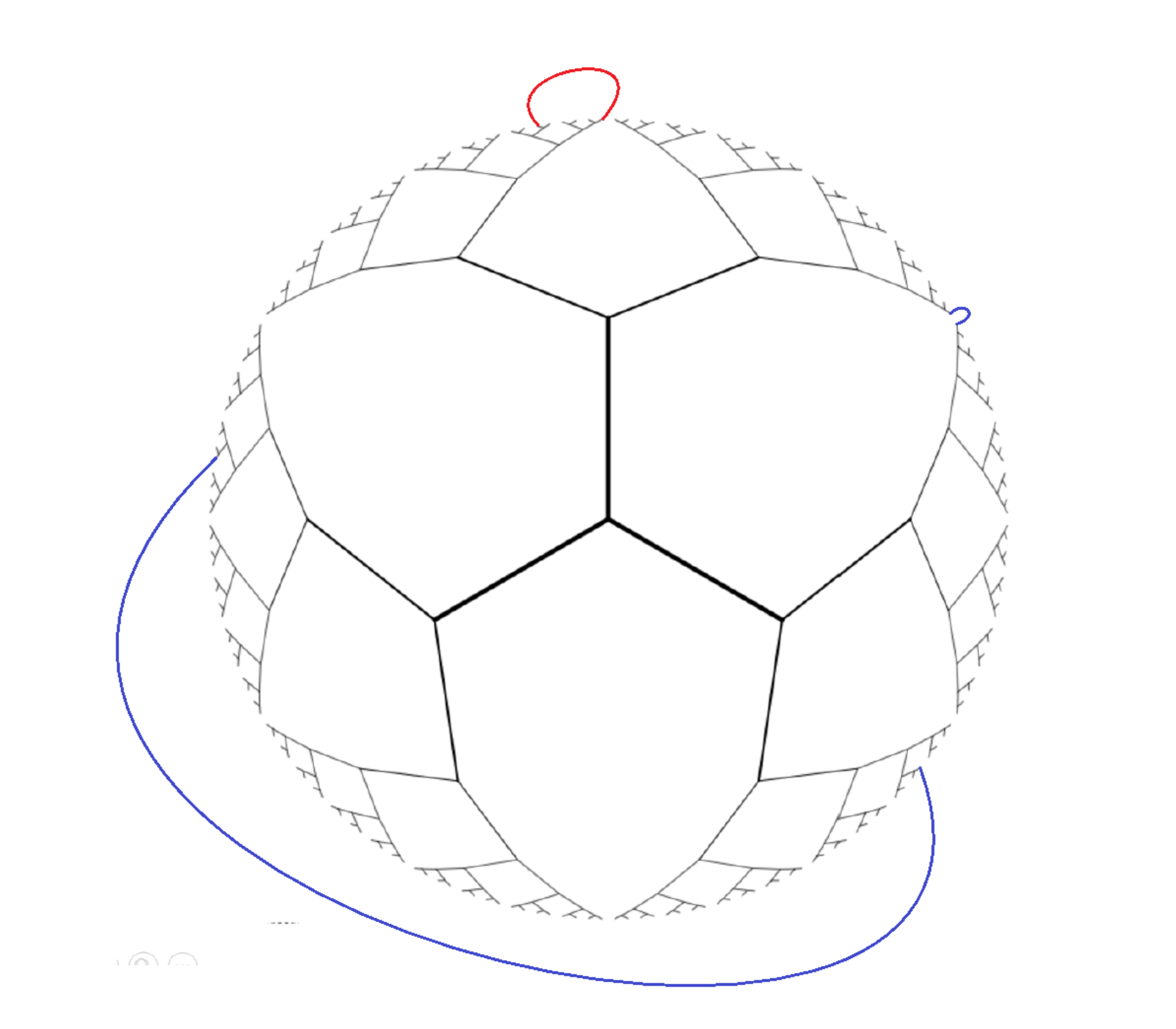}
\caption{Graph ceases being a tree and doubles back. Since group is homogeneous the structure must look the same from every vertex. The shortest loops (girth of graph) is of order $4^K.$ Thus the red loop is too short. The loops must look more like the blue loop.}
\label{bethe-2}
\end{center}
\end{figure}

\bn
But once we reach $\CC_{max}$ the graph can not continue to grow. Collisions occur and loops must form. The graph must double back on itself and revisit previously visited epsilon-balls\footnote{If we follow an epsilon ball from the identity it will not in general perfectly coincide with an epsilon ball after executing a loop. There is a bit of sloppiness but it doesn't seem to be important. Note that if $\epsilon $  is decreased the maximum complexity increases and the graph becomes bigger.}. We show a couple of possible loops that might form in figure \ref{bethe-2}.
Now we know that any loop which passes through the central node must be very big, namely of length $\sim 4^K.$  But because the space $\suk$ is a group space, every point on the graph is the same as every other point. Thus it must be that loops passing through any point must be equally long. It is clear from the figure that loop containing the red segment is much shorter and should not occur, but the loop containing the blue segment is long and may occur.

\bn
Let me summarize the conjectured properties of circuit graphs generated by iterating one-step circuits: 

\begin{enumerate}
\item The degree is the same for all vertices and is given by
\be 
\rm degree \it = \frac{K!}{\left(  K/2 \right)!}  \sim \left(  \frac{2K}{e}   \right)^{\frac{K}{2}}.
\ee
This is much smaller than the number of vertices. The graph can be said to be sparse.
\item The number of vertices in the graph is of order $e^{4^K}.$
\item  The greatest distance between vertices (diameter)  is $\CC_{max} \sim 4^K$. The diameter is therefore logarithmic in the number of vertices.
\item Loops of length less than $4^K$ are rare or absent.
\item The graph is homogenous and from any point looks tree-like out to distances of order the diameter.
\end{enumerate}

These properties are very familiar to graph theorists. They are the properties of a good \it expander \rm graph. I think it may be even stronger. It may be among the strongest expanders called Ramanujan graphs. Graphs of this type are discrete analogs of finiite negatively curved geometries such as the hyperbolic plane, with identifications that render it compact\footnote{The relation between complexity and such geometries was described in Brown, Susskind and Zhao, [arXiv:1608.02612 [hep-th]].}

Quantum complexity has a strong geometrical aspect that was first appreciated by Nielsen. I would make the case that it 
 is basically the subject of right-invariant geometries on a group space with the parameters chosen so that the curvature is negative.

\begin{figure}[H]
\begin{center}
\includegraphics[scale=.2]{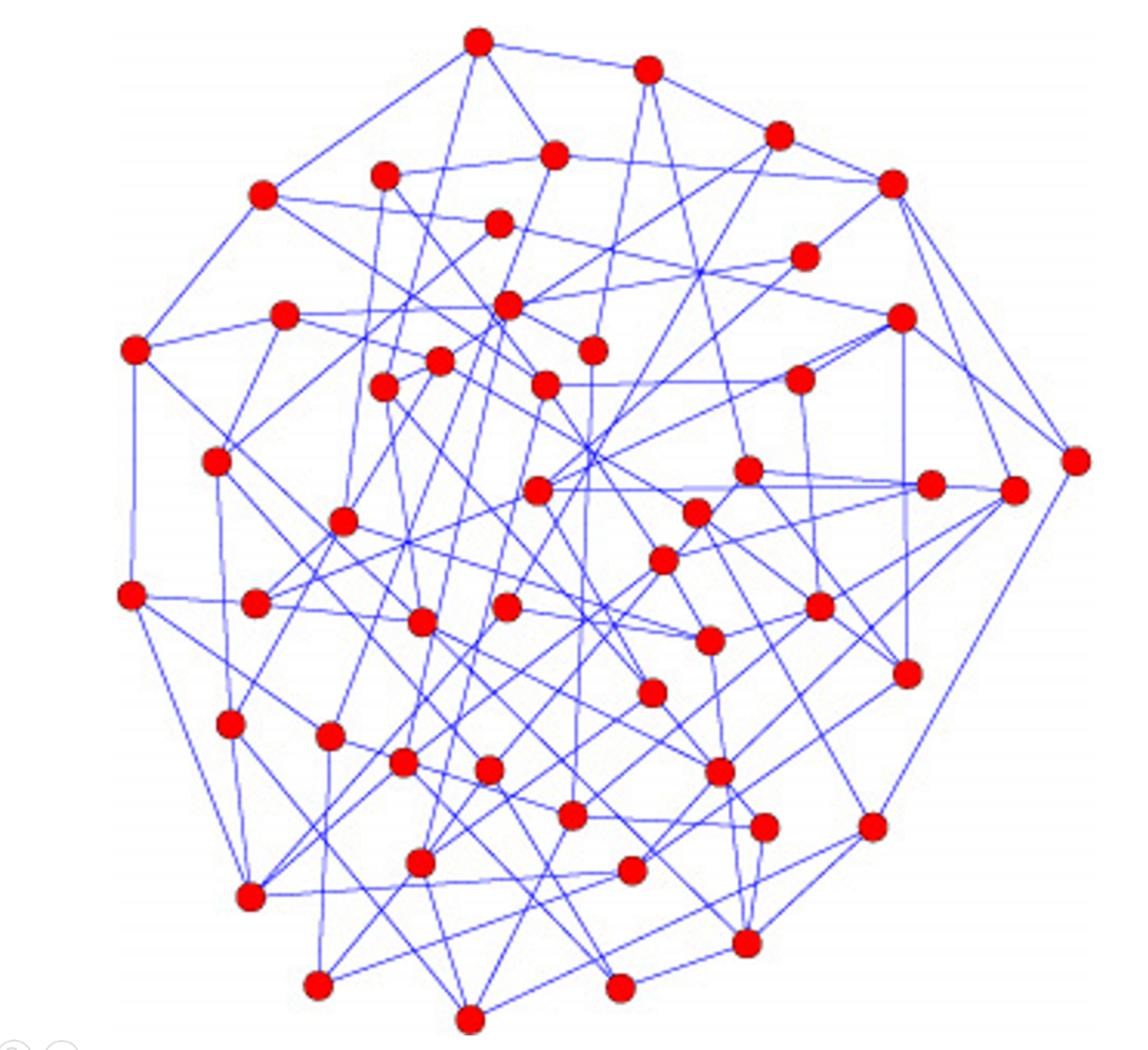}
\caption{50-vertex, (almost) regular $d=4$ expander. Generated by some AI optimization  protocol.} 
\label{expander}
\end{center}
\end{figure}
\begin{figure}[H]
\begin{center}
\includegraphics[scale=.2]{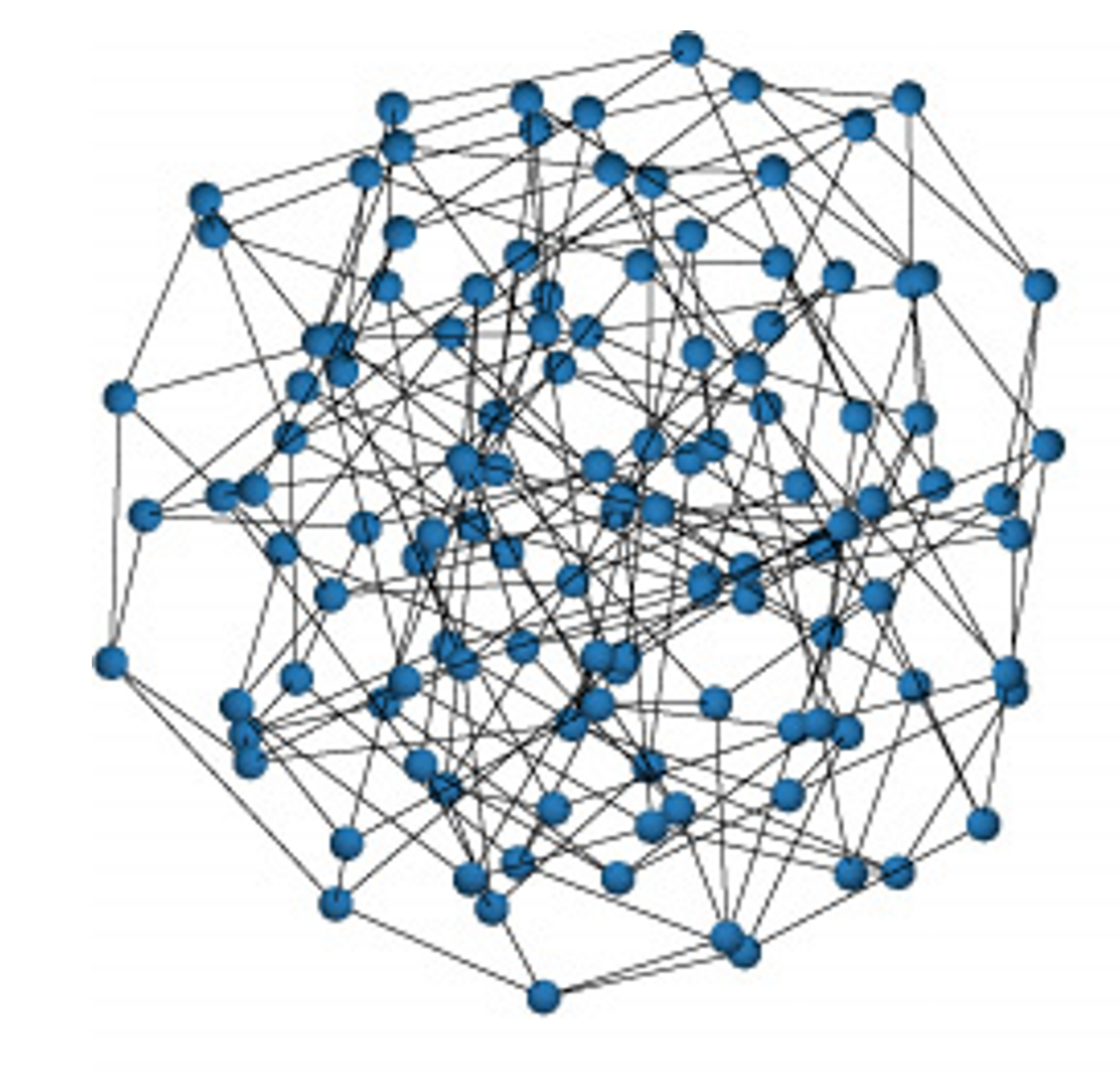}
\caption{120 vertex, regular  $d=4$ Ramanujan-expander. Quoting Donetti, et al.:
``Some remarks on its topological properties are in order. The average distance 3.714 is relatively small (i.e. one
reaches any of the 120 nodes starting from any arbitrary origin in less than four steps, on average."  Note: $\log{120}=4.78749174278$. Not too different from  3.714.  The authors then go on to say, ``The clustering coefficient vanishes,
reflecting the absence or short loops (triangles), and the minimum loop size is large, equal to 6, and identical for
all the nodes. In a nutshell: the network homogeneity is remarkable; all nodes look alike, forming rather intricate,
decentralized structure, with delta-peak distributed topological properties."}   
\label{expander2}
\end{center}
\end{figure}

\bn

\section{The Second Law of Quantum Complexity}

 I'm not sure why complexity theorists never remarked on the similarity of quantum complexity and classical entropy, or the existence of a second law of complexity\footnote{There is a very interesting paper by Zurek  in which he argues that that the classical entopy of a system is the ensemble average of the Kolmogorov algorithmic complexity of states in phase space. [W. Zurek,  Phys.\ Rev.\ A {\bf 40}, 4731 (1989)] The idea is very similar to the connection between quantum-computational-complexity and entropy described by Brown et.al in   [arXiv:1701.01107 [hep-th]], and described here. }. The second law  is the subject of  lecture three but I will briefly explain it here because it plays an important role in lecture two.
 Just to be clear, although ensemble-averaged complexity is a kind of entropy it is NOT conventional entropy. The conventional entropy of a system of $K$ qubits is bounded by $K\log{2}$, the logarithm of the maximum number of mutually orthogonal vectors in the Hilbert space. By contrast the maximum  complexity is exponential in $K$.  Maximum complexity $\sim 4^K\log{1/\epsilon}$, the number of $\epsilon$-balls in $\suk.$ 

The quantity $4^K \log{}1/\epsilon$ 
does have an interpretation in terms of the entropy of a classical \it auxiliary \rm system 
$\CA$ associated with the quantum system of $K$ qubits. We may think of it as the maximum entropy of a classical system with $4^K$ classical degrees of freedom.  The auxiliary  system is just the classical $4^K$ collection of coordinates\footnote{Stictly speaking, $4^K-1$.} that describe the evolving time-evolution operator $U(t).$ Figure \ref{motion} illustrates the auxiliary system. In lecture three this will be made more precise. Quantities describing the auxiliary system will carry a subscript $\CA.$

For now let us consider an simplified version of the quantum evolution of a system at high temperature. We envision an
ensemble of fictitious particles moving on $\suk$. The particles are  random walkers which all start at the origin of $\suk$ (the identitiy operator) at some initial time. The dynamics is discrete: at each step the position is updated by applying a depth-one circuit. This means the particles execute random walks on the graph
that I just explained.

At each vertex the decision for the next step is made randomly. I will allow the possibility of back-steps along the previous edge. 
Initially the probability is concentrated at the origin and the entropy of the fictitious system is zero.

After one time-step the particle is at the first level of the tree on one of the leaves, as in figure \ref{spider}. The fictitious entropy is $\log{d}$ and the complexity is $\frac{K}{2}.$ In the next step the particle has a probability $\sim 1/d$ to back-track, but for large $d$ that is negligible. With probability close to one the particle moves outward to the next level where the complexity is 
$K.$ and the auxiliary  entropy $S_{\CA}$ is $\log{d^2} = 2\log{d}.$
\bn

After $n$ steps the particle with high probability is at the $n^{th}$ level, the auxiliary entropy is $S_{\CA} = n\log{d}$,  and the complexity is
$nK/2.$
Evidently the complexity and fictitious entropy of the auxiliary system  are related,
\be 
S_{\CA} \approx  \CC  \  \log{K}
\ee

This identification is dependent on the negative curvature and high dimensionality of complexity space. These two ingredients are what insure that collisions are rare and that we can identify the level of the tree with the minimum distance from the origin at $I$. In other words, up to a factor $K/2$ we may identify the depth with complexity.

Of course it is not rigorously true that there are no collisions. It's just that collisions are rare for sub-exponential time. We can be reasonably sure that almost all leaves  (vertices) have complexity proportional to their level, but it is much harder to know that a given leaf has not had collisions in the  past. It's for this reason that we identify the entropy $S_{\CA}$ with the ensemble averaged complexity.

The second law of complexity is just the second law of thermodynamics---the overwhelming statistical likelihood that entropy will increase---applied to the ensemble average of complexity. The reason why complexity almost always increases when it is less than maximum is the same as why classical entropy  almost always increases when it is less than maximum---the number of states exponentially increases with increasing entropy/complexity.

Let us follow a particular member of the ensemble. As long as it is not an exponential number of steps from $I$ the complexity will simply reflect the exponentially growing number of states as we mover outward from the origin. It will with very high probability increase linearly with time. 
However once $\CC \sim 4^K$ the particle will have reached the maximum distance on the graph and the complexity will stop increasing. Complexity equilibrium will have been achieved. The number of states with  maximum complexity is so vast, that the particles will get lost among them and remain at maximum complexity for a recurrence time. The recurrence time for the classical system will be $t_{recur} = \exp{S_{\CA}}$ which is doubly exponential in $K,$.
\be
t_{recur} \sim e^{4^K}
\ee

Thus we expect a singly  exponential time $\sim 4^K$ during which complexity linearly grows, after which it remains approximately constant at its maximum. But then, on gigantically long time scales $\sim \exp\exp K$ it will recur to small values, and $U(t)$ will return to the neighborhood of the identity.

\begin{figure}[H]
\begin{center}
\includegraphics[scale=.3]{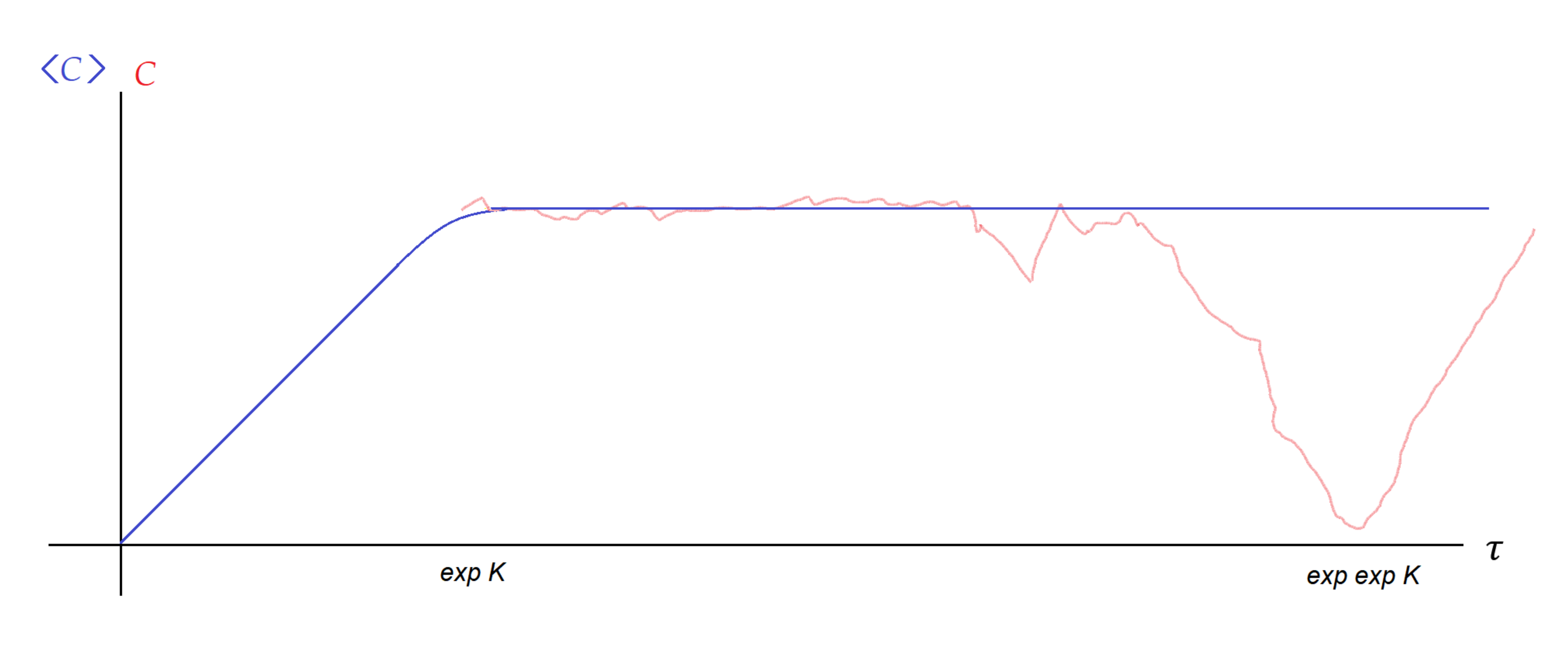}
\caption{Evolution of complexity with time. The ragged red curve is the evolution for a specific instance of an ensemble. The smooth curve is the ensemble average.}
\label{2nd-law}
\end{center}
\end{figure}

The transition from linear growth to complexity equilibrium is very sudden.

\begin{figure}[H]
\begin{center}
\includegraphics[scale=.2]{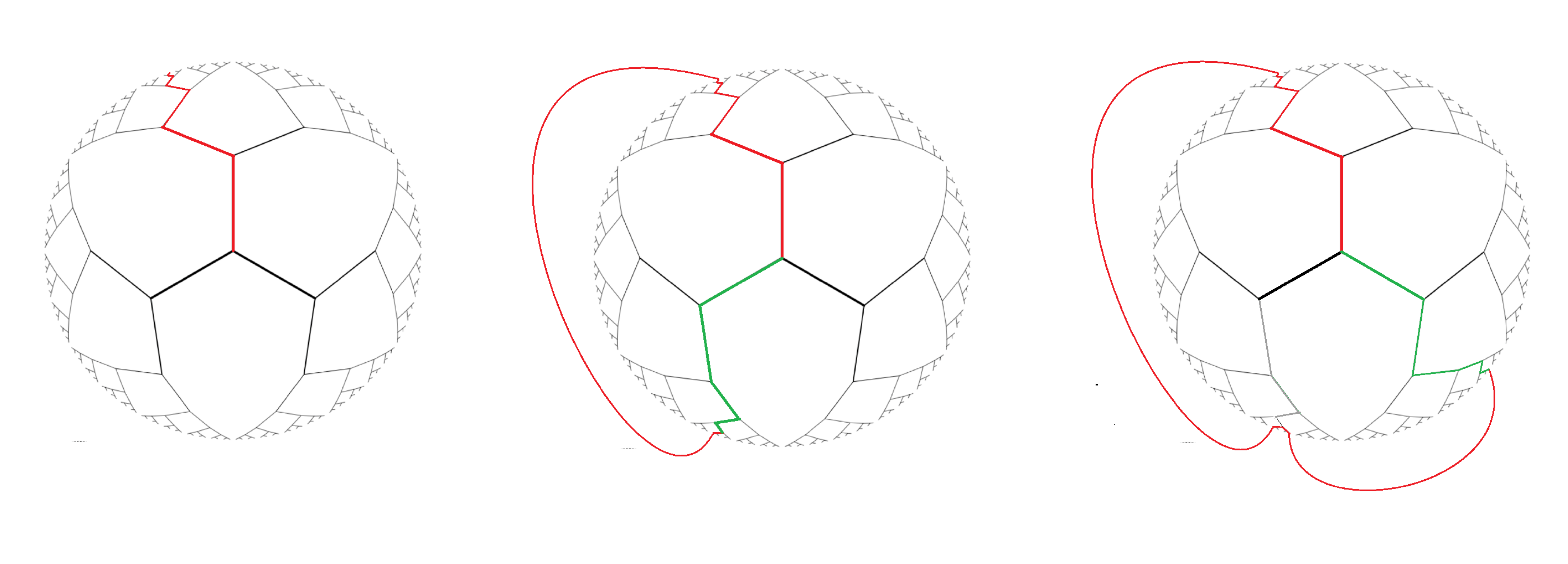}
\caption{The left panel shows a (red) trajectory beginning at the identity and proceeding outward along toward greater complexity. It eventually reaches maximum complexity. The middle panel follows the evolution as it doubles back to a slightly less-than-maximal complexity. At this point the minimal path jumps to the green path. In the third panel the trajectory continues to maximal complexity and then jumps again. }
\label{cut}
\end{center}
\end{figure}
\bn
It is interesting to see how this behavior can be understood in terms of circuit graphs. We can get an idea by looking at figure \ref{cut}. The figure shows the evolution of the $\CA$ auxiliary system moving according to some ``dynamical" rule that mocks up Hamiltonian evolution starting with $U(0) = I.$ The trajectory moves outward and until collisions occur the complexity increases linearly with time. The complexity is proportional to the graph distance from $I$. That's shown as  the red trajectory in the left panel.

Once the trajectory reaches maximum complexity it keeps going, but it has no choice but to visit previously visited sites. With overwhelming probability it will jump to another almost maximally complex state. That is shown in the middle panel. 

But now, there is a shorter path to the end of the red trajectory. It is shown in green. The complexity is \bf NOT \rm the length of the red trajectory but rather the length of the green trajectory. Thus the complexity does not increase and may even decrease a bit.

From there the dynamical red trajectory continues. With overwhelming likelihood it moves outward because the overwhelming number of branches reach outward. It soon reaches the next point where it has to loop around. A new green trajectory forms that is quite different than the previous one, but also has close to maximal complexity. That is the reason why the top of the curve in figure \ref{2nd-law} is a bit ragged.

Eventually the auxiliary particle will find its way back to low complexity and the cycle will repeat but this takes a quantum-recurrence time.

Note that the transition from linearly increasing complexity to equilibrium at the top of the curve is sudden. It's similar to a first order phase transition where another local minimum crosses over and becomes the global minimum.

\bn

That concludes lecture I. In the next lecture I will explain what   the complexity and its second law has to do with black holes.

\end{document}